\documentclass[aps,pra,twocolumn,superscriptaddress]{revtex4}

\usepackage[latin1]{inputenc}
\usepackage[T1]{fontenc}
\usepackage{amsmath,amssymb}
\usepackage{bbm}
\usepackage[squaren]{SIunits}
\usepackage{units}
\usepackage[dvips]{graphicx}

\DeclareMathOperator{\diag}{diag}
\DeclareMathOperator{\var}{Var}

\begin{document}


\title{Magnetometry with entangled atomic samples}

\author{Vivi Petersen}
\affiliation{QUANTOP - Danish National Research Foundation Center for
  Quantum Optics}
\affiliation{Department of Physics and Astronomy, University of
  Aarhus, DK-8000 Århus C, Denmark}
\author{Lars Bojer Madsen}
\affiliation{Department of Physics and Astronomy, University of
  Aarhus, DK-8000 Århus C, Denmark}
\author{Klaus Mølmer}
\affiliation{QUANTOP - Danish National Research Foundation Center for
  Quantum Optics}
\affiliation{Department of Physics and Astronomy, University of
  Aarhus, DK-8000 Århus C, Denmark}

\date{\today}

\begin{abstract}
We present a theory for the estimation of a scalar or a vector
magnetic field by its influence on an ensemble of trapped spin
polarized atoms. The atoms interact off-resonantly with a
continuous laser field, and the measurement of the polarization
rotation of the probe light, induced by the dispersive atom-light
coupling, leads to spin-squeezing of the atomic sample which
enables an estimate of the magnetic field which is more precise
than that expected from standard counting statistics. For
polarized light and polarized atoms, a description of the
non-classical components of the collective spin angular momentum
for the atoms and the collective Stokes vectors of the light-field
in terms of effective gaussian position and momentum variables is
practically exact. The gaussian formalism describes the dynamics
of the system very effectively and accounts explicitly for the
back-action on the atoms due to measurement and for the estimate
of the magnetic field. Multi-component magnetic fields are
estimated by the measurement of suitably chosen atomic observables
and precision and efficiency is gained by dividing the atomic gas
in two or more samples which are entangled by the dispersive
atom-light interaction.
\end{abstract}

\pacs{}

\maketitle

\section{Introduction}
\label{sec:introduction}

Precision atomic magnetometry relies on the measurement of the Larmor
precession of a spin-polarized atomic sample in a magnetic field
\cite{budker02,kominis03,auzinsh04:_can_a}. From standard counting
statistics arguments, one might expect the uncertainty in such
measurements to decrease with the interaction time $t$ and with the
number of atoms $N_\text{at}$ as $1/\sqrt{N_\text{at} t}$. If, on the
other hand, the monitoring of the atomic sample, necessary for the
read-out of the estimate of the magnetic field, squeezes the atomic
spin, the above limit may be surpassed.  In a recent theoretical
analysis it was considered to estimate a scalar $B$ field by a
polarization rotation measurement of an off-resonant light beam
passing through a trapped cloud of spin-1/2 atoms.  This interaction
squeezes the spin of the atomic sample, and by quantum trajectory
theory~\cite{carmichael93:_open_system_approac_quant_optic} combined
with the classical theory of Kalman
filters~\cite{geramia03:_quant_kalman_filter_heisen_limit_atomic_magnet,stockton04},
the uncertainty in the field strength was found to decrease as
$1/(N_\text{at}
t^{3/2})$~\cite{geramia03:_quant_kalman_filter_heisen_limit_atomic_magnet}.
Very recently this proposal was implemented experimentally, and indeed
sub-shotnoise sensitivity was found~\cite{geremia04:_sub-shotnoise}.

In a recent analysis of the experiment, we advocated treating all
variables, including the magnetic field, as quantum
variables~\cite{molmer04:_estim_class_param_gauss_probes}. Secondly,
we introduced a gaussian approximation at an early stage in the
formulation of the theory. We further motivated and developed this
point of view in a detailed discussion of the spin-squeezing
process~\cite{madsen04:_spin_squeez_precis_probin_light}. As mentioned
in these works, the gaussian approximation is essentially exact for
the atomic and photonic degrees of freedom of the system under
concern, and the advantages obtained by introducing this description
from the outset of the theoretical treatment are at least four-fold:
(i) the gaussian description explicitly accounts for the dynamics of
the system \textit{and} its behavior under measurements through update
formulae for the expectation values and the covariance matrix which
together fully characterize the gaussian state, (ii) the numerical
treatment of the update formulae involves only the manipulation of
low-dimensional matrices, (iii) in the limit of small time-steps the
update formula for the covariance matrix translates into a matrix
Ricatti differential equation which often lends itself to analytical
solution, and (iv) effects of noise introduced by, e.g., photon
absorption and atomic decay are readily included.

Here, we extend our previous
analysis~\cite{molmer04:_estim_class_param_gauss_probes} to explore
the possibilities for estimating $B$ fields with not only one, but
also two or three spatial components. In cases with more than one
component, it is advantageous to use two or more polarized atomic
samples. With such setups, we may identify sets of commuting
observables which allow a simultaneous estimate of the $B$ field
components. We also discuss how to gain precision and efficiency by
entangling the atomic gasses.

The paper is organized as follows. In Sec.~\ref{sec:at-ph-probe}, we
introduce the atom-photon system used for the estimation of the
magnetic field and describe the atomic and photonic gaussian
variables. In Sec.~\ref{sec:meas-one-magn}, we investigate the
estimation of a single $B$ field component, we describe our theoretical
method in some detail and we derive an analytical solution for the
decrease in variance of the $B$ field as a function of time. In
Sec.~\ref{sec:meas-two-magn}, we present our results for the
estimation of two or three spatial $B$ field components. In
Sec.~\ref{sec:entanglement}, we quantify the entanglement between the
samples used in our optimal protocol for the estimation of several $B$
field components. In Sec.~\ref{sec:noise}, we explain how to include
noise in the description and we study the effects of noise on the
precision of measurements. In Sec.~\ref{sec:conclusion-outlook}, we
conclude and present an outlook.

\section{Atom-light system: collective variables}
\label{sec:at-ph-probe} 

To estimate the strength of a $B$ field,
we let it interact with an atomic spin-system which is
continuously probed by a light beam along the lines of
Refs.~\cite{kuzmich98:_atomic_quant_non_demol_measur_squeez,takahasi99,kuzmich99,
duan00:_quant_commun_atomic_ensem_using_coher_light,
julsgaard01:_exper_long_entan_two_macros_objec,
molmer04:_estim_class_param_gauss_probes,madsen04:_spin_squeez_precis_probin_light}.
In short, we imagine to have a gas of trapped spin-1/2 atoms which
are described by a collective spin operator $\mathbf{J} =
\frac{\hbar}{2}\sum_i \boldsymbol{\sigma}_i$ with
$\boldsymbol{\sigma}_i$ the Pauli spin matrices. The atoms are
initially pumped such that they are polarized along the $x$ axis
and $J_x$ can be treated as a classical variable $\langle
J_x\rangle = \frac{\hbar
  N_\mathrm{at}}{2}$ with $N_\mathrm{at}$ the number of
  atoms. The other two projections of the spin, $J_y$
  and $J_z$, obey the commutation relation $[ J_y, J_z] = i \hbar
  J_x$ which may be rewritten as $[x_\text{at}, p_\text{at}] = i$
for the effective position and momentum variables
$  x_\mathrm{at} = \frac{J_y}{\sqrt{\hbar\langle J_x\rangle}},
  p_\mathrm{at} = \frac{J_z}{\sqrt{\hbar\langle J_x\rangle}}.$
The uncertainty is easily shown to be minimal in the initial state
and, hence, the state pertaining to $x_\text{at}$ and
$p_\text{at}$ is gaussian.

The light beam propagates along the $y$ axis and is linearly
polarized along $x$ such that its Stokes operator $\langle
S_x\rangle = \frac{\hbar N_\mathrm{ph}}{2}$ is classical with
$N_\mathrm{ph}$ the number of photons. The two remaining
components fulfill a commutator relation similar to the atomic spin
case. Accordingly, for the effective variables
$  x_\mathrm{ph} = \frac{S_y}{\sqrt{\hbar\langle S_x\rangle}},
  p_\mathrm{ph} = \frac{S_z}{\sqrt{\hbar\langle S_x\rangle}}$,
we have $[x_\text{ph}, p_\text{ph}] = i$ and the initial coherent
state of the field is a minimum uncertainty gaussian state.

As shown in Ref.~\cite{madsen04:_spin_squeez_precis_probin_light}
and references therein, the light and atomic variables evolve as
\begin{align}
\label{eq:evo1} x_\text{at} &\mapsto x_\text{at} + \kappa_\tau
p_\text{ph}, &p_\text{at} \mapsto p_\text{at} \\
\label{eq:evo2}
 x_\text{ph} &\mapsto \kappa_\tau p_\text{at} +
x_\text{ph}, & p_\text{ph} \mapsto p_\text{ph}
\end{align}
when a segment of probe light of duration $\tau$ and flux $\Phi$, and
with a characteristic atom-light coupling $\kappa_\tau \propto
\sqrt{\langle J_x\rangle \Phi\tau}$, is transmitted through the gas.
The present dynamics in combination with a detection of the
$x_\text{ph}$ component of the photon field leads to squeezing of the
$p_\text{at}$ component of the atomic spin.

\section{Estimating One spatial component of a Magnetic Field}
\label{sec:meas-one-magn} 

\begin{figure}[htbp]
  \centering
  \includegraphics{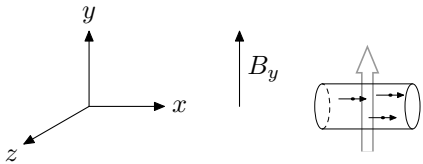}
  \caption{Setup for measuring the $B$ field component along the
    $y$ axis. This is done using one atomic gas polarized along the
    $x$ axis and one photon probe beam propagating along the $y$ axis with a
    classical $S_x$.}
  \label{fig:5}
\end{figure}

The problem of estimating a single component of a $B$ field was
treated in Ref.~\cite{molmer04:_estim_class_param_gauss_probes} and
only a brief discussion is included here for completeness.
Figure~\ref{fig:5} shows the setup.
The $B$ field directed along the $y$ direction, causes a Larmor
rotation of the atomic spin towards the $z$ axis, i.e., in time
$\tau$, $p_\text{at}$ evolves as $p_\text{at} \mapsto p_\text{at} -
\mu_\tau B$ where $\mu_\tau$ is given by the magnetic moment $\beta$,
via $\mu_\tau = \frac{1}{\hbar}\beta \sqrt{\frac{\langle
    J_x\rangle}{\hbar}}
\tau$~\cite{molmer04:_estim_class_param_gauss_probes}. Hence,
Eqs.~\eqref{eq:evo1}--\eqref{eq:evo2}, generalize to
\begin{equation}
 \label{eq:pat-B}
 \mathbf{y} \mapsto \mathbf{S}_\tau\mathbf{y}
\end{equation}
with $\mathbf{y} = (B_y, x_\mathrm{at}, p_\mathrm{at},
x_\mathrm{ph}, p_\mathrm{ph})^T$ and
\begin{gather}
  \label{eq:12}
  \mathbf{S}_\tau =
  \begin{pmatrix}
    1 & 0 & 0 & 0 & 0\\
    0 & 1 & 0 & 0 & \kappa_\tau\\
    -\mu_\tau & 0 & 1 & 0 & 0\\
    0 & 0 & \kappa_\tau & 1 & 0\\
    0 & 0 & 0 & 0 & 1
  \end{pmatrix}.
\end{gather}
It is the coupling of the $B$ field to the spin-squeezed variable
$p_\text{at}$ that makes an improved precision measurement of the
magnetic field
possible~\cite{geramia03:_quant_kalman_filter_heisen_limit_atomic_magnet}.

Gaussian variables have been studied widely in relation to
entanglement~\cite{giedke03:_entan_trans_pure_gauss_states}. The
gaussian description can be used as long as the interactions in the
system are at most second order polynomials in the position and
momentum operators and only homodyne measurements are carried out on
the
observables~\cite{giedke02:_charac_gauss_operat_distil_gauss_states}.
In particular, the operations of relevance for this work preserve the
character of a gaussian state. We treat the continuous probe beam as a
succession of small beam segments of duration $\tau$, and let the
state develop during time $\tau$ between successive measurements. We
recall that a gaussian state is fully characterized by its mean value
vector $\mathbf{m} = \langle \mathbf{y} \rangle$ and its covariance
matrix $\boldsymbol{\gamma}$ where $\gamma_{ij} = 2\mathrm{Re}
\langle(y_i-\langle y_i\rangle) (y_j-\langle y_j\rangle)\rangle$.
Accordingly, we only need update formulae for $ \mathbf{m}$ and
$\boldsymbol{\gamma}$. The initial covariance matrix is $\gamma_0 =
\diag(2\var(B_0), 1, 1, 1, 1)$ with $\var(B_0)$ the initial variance
of the $B$ field. Under the linear transformation~\eqref{eq:pat-B},
$\mathbf{m}$ and $\boldsymbol{\gamma}$ transform as
\begin{align}
  \label{eq:11}
  \mathbf{m}(t+\tau) &= \mathbf{S}_\tau \mathbf{m}(t)\\
  \label{eq:28}
  \boldsymbol{\gamma}(t+\tau) &= \mathbf{S}_\tau \boldsymbol{\gamma}(t)
  \mathbf{S}_\tau^T.
\end{align}

The photon field is monitored continuously by  detection of
$x_\text{at}$. A major advantage of the gaussian description is
that  the back-action on the residual system due to measurement is
explicitly given. We write the covariance matrix as~\cite{fiurasek02:_gauss_trans_distil_entan_gauss_states,giedke02:_charac_gauss_operat_distil_gauss_states,eisert03:_introd_basic_entan_theor_contin_variab_system}
\begin{gather}
  \label{eq:13}
  \boldsymbol{\gamma} =
  \begin{pmatrix}
    \mathbf{A}_\gamma & \mathbf{C}_\gamma\\
    \mathbf{C}_\gamma^T & \mathbf{B}_\gamma
  \end{pmatrix},
\end{gather}
with $\mathbf{A}_\gamma$ the covariance matrix for the $B$ field
and atoms, $\mathbf{y}_1 = (B_y, x_\mathrm{at}, p_\mathrm{at})^T$,
$\mathbf{B}_\gamma$ the covariance matrix for the
\begin{figure}[htbp]
  \centering
  \includegraphics[width=7cm]{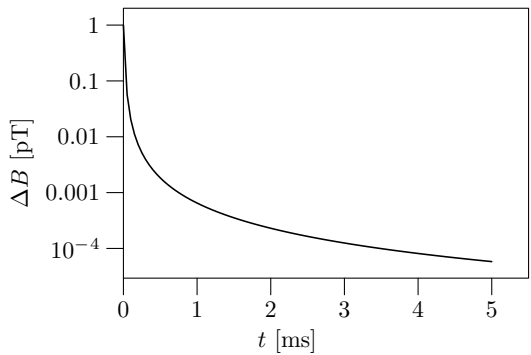}
  \caption{Uncertainty of one $B$ field component as a function of
    time. The value at $t={\unit[5]{\milli\second}}$ is $\Delta B_y =
    {\unit[5.814\times 10^{-5}]{\pico\tesla}}$. We have chosen a
    segment duration $\tau = 10^{-8}$~s and corresponding field
    parameters $\kappa_\tau^2 = 0.0183$ and $\mu_\tau = 8.8 \times
    10^{-4}$.}
  \label{fig:1}
\end{figure}
photons, $\mathbf{y}_2 = (x_\mathrm{ph}, p_\mathrm{ph})^T$, and
$\mathbf{C}_\gamma$ the correlation matrix for $\mathbf{y}_1$ and
$\mathbf{y}_2^T$. The measurement of $x_\mathrm{ph}$ then
transforms these matrices according
to~\cite{fiurasek02:_gauss_trans_distil_entan_gauss_states,eisert03:_introd_basic_entan_theor_contin_variab_system}
\begin{subequations}
  \label{eq:27}
  \begin{align}
    \label{eq:14}
    \mathbf{A}_\gamma &\mapsto \mathbf{A}_\gamma -
    \mathbf{C}_\gamma(\pi\mathbf{B}_\gamma\pi)^{-}\mathbf{C}_\gamma^T,\\
    \label{eq:15}
    \mathbf{B}_\gamma &\mapsto \mathbbm{1}_{2\times2},\\
    \label{eq:16}
    \mathbf{C}_\gamma &\mapsto 0,
  \end{align}
\end{subequations}
where $\pi = \diag(1, 0)$, and $()^-$ denotes the
Moore-Penrose pseudo-inverse. Equations~\eqref{eq:15} and
\eqref{eq:16} follow from the fact that a new light segment is used
in every measurement: The initial covariance matrix for every new
segment of a coherent photon beam is the identity matrix, and
immediately after measurement there are no correlations between
$\mathbf{y}_1$ and $\mathbf{y}_2$ so $\mathbf{C}_\gamma$ is
substituted with the zero matrix.

The time evolution of $\mathbf m$ depends on the actual
measurements in the optical detection which is a random process.
Therefore the  time evolution of the mean value vector is a
stochastic process, and it transforms
as~\cite{fiurasek02:_gauss_trans_distil_entan_gauss_states,giedke02:_charac_gauss_operat_distil_gauss_states,molmer04:_estim_class_param_gauss_probes}
\begin{gather}
  \label{eq:29}
  \mathbf{m}_1 \mapsto \mathbf{m}_1 + \mathbf{C}_\gamma
  (\pi\mathbf{B}_\gamma\pi)^{-} (\chi,\cdot)^T,
\end{gather}
where $\chi$ is the difference between the measurement outcome and the
expectation value of $x_\mathrm{ph}$, i.e., a gaussian random variable
with mean value zero and variance $\nicefrac{1}{2}$. Since $
(\pi\mathbf{B}_\gamma\pi)^{-} = \diag(B_{11}^{-1},0)$ with $B_{11}$
twice the variance of $x_\text{ph}$, the second entrance in the vector
$(\chi, \cdot)$ need not be specified. In closing this section, we
note that it is possible to understand the transformations in
Eqs.~\eqref{eq:14} and \eqref{eq:29} by the corresponding
transformation of classical gaussian probability
distributions~\cite{madsen04:_spin_squeez_precis_probin_light,maybeck79:_stoch_model_estim_contr}.

\subsection{Analytical solution}
\label{sec:analytical-solution}

From the update formulae, we may, in the limit of small time
increments, derive a differential equation for
$\mathbf{A}_\gamma$. As $B_y$ only causes rotation perpendicular
to its direction, the variable $x_\mathrm{at} \propto J_y$ does
not couple to ($B_y, p_\mathrm{at}$) and, hence, we only need to
consider a $2\times2$ system with $\mathbf{y} = (B_y,
p_\mathrm{at})^T$. The pertaining differential equation is on the
matrix Ricatti form~\cite{stockton03:_robus_quant_param_estim}
\begin{gather}
  \label{eq:20}
  \dot{\mathbf{A}}_\gamma(t) = \mathbf{C} -
  \mathbf{D}\mathbf{A}_\gamma(t) - \mathbf{A}_\gamma(t)\mathbf{E} -
  \mathbf{A}_\gamma(t)\mathbf{B}\mathbf{A}_\gamma(t),
\end{gather}
with $\mathbf{C} = 0$, $\mathbf{D} = \left(
\begin{smallmatrix}
  0 & 0\\
  \mu & 0
\end{smallmatrix}
\right)$, $\mathbf{E} = \mathbf{D}^T$, and $\mathbf{B} = \left(
\begin{smallmatrix}
  0 & 0\\
  0 & \kappa^2
\end{smallmatrix}
\right)$ where $\kappa^2 = \kappa_\tau^2/\tau$ and $\mu = \mu_\tau/\tau$.
As may be checked by insertion, the solution to Eq.~\eqref{eq:20} is
$\mathbf{A}_\gamma = \mathbf{W}\mathbf{U}^{-1}$, where
$\dot{\mathbf{W}} = -\mathbf{D}\mathbf{W} + \mathbf{C}\mathbf{U}$ and
$\dot{\mathbf{U}} = \mathbf{B}\mathbf{W} + \mathbf{E}\mathbf{U}$. The
linear differential equations for $\mathbf{W},\mathbf{U}$ can be
solved, and the resulting solution for the variance of the $B$ field
reads:
\begin{gather}
  \label{eq:21}
  \begin{split}
    \var(B(t)) &= \frac{\var(B_{0}) (\kappa^2t+1)}{
      \frac{1}{6}\kappa^4\mu^2\var(B_{0})t^4 +
      \frac{2}{3}\kappa^2\mu^2\var(B_{0})t^3 +
      \kappa^2t + 1}\\
    &\xrightarrow[t\to\infty]{} \frac{6}{\kappa^2\mu^2t^3} \propto
    \frac{1}{N_\mathrm{at}^2 \Phi t^3},
  \end{split}
\end{gather}
where we have introduced the photon flux $\Phi$.

Figure~\ref{fig:1} shows the decrease in the uncertainty of the $B$
field with time in a calculation with physically realizable
parameters. From standard counting statistics one might have expected
the variance to decrease as $\nicefrac{1}{N_\mathrm{at}}$. Due to the
measurement-induced squeezing of the atomic spin, we do, however,
obtain the faster $\nicefrac{1}{N_\mathrm{at}^2}$ decrease. As we
shall discuss in Sec.~\ref{sec:noise}, the inclusion of noise due to
decoherence of atomic spins will alter this dependence on time and on
the number of atoms.

\section{Estimating two or three spatial components of a magnetic field}
\label{sec:meas-two-magn}

In this section, we describe how to estimate two or three spatial
components of a $B$ field. In the setup in Fig.~\ref{fig:5}, we
obtained an estimate of one component by using one atomic gas and
one probe beam. To estimate more components we shall need more
probe beams and more atomic gasses.

In order to make a fair comparison of different schemes, we shall
assume that all measurements are carried out in a time interval of the
same duration, e.g., ${\unit[5]{\milli\second}}$ as in
Fig.~\ref{fig:1}, and that the total photon number used and the total
number of atoms are kept constant.

\subsection{Two components: two probe beams and one or two separate
  atomic gasses}
\label{sec:two-probe-beams}

In order to estimate $B_z$ in addition to $B_y$, we observe that $B_z$
causes a rotation of $x_\mathrm{at} \propto J_y$ so if we add a second
optical probe beam propagating along $z$ in Fig.~\ref{fig:5} and with
$S_x$ classical then $x_\mathrm{at}$ will cause a rotation of the
field variable $p_\mathrm{ph} \propto S_z$ on the second beam which we
can then measure. The setup is symmetric with respect to $y$
and $z$, and we obtain equal uncertainties on $B_y$ and $B_z$. The
problem with this approach is that unlike the unknown classical
quantities $B_y$ and $B_z$, the atomic observables $x_\mathrm{at}$ and
$p_\mathrm{at}$ do not commute. While the first beam squeezes
$x_\mathrm{at}$, it anti-squeezes $p_\mathrm{at}$, and the other beam
does the opposite.
\begin{figure}[htbp]
  \centering
  \includegraphics[width=7cm]{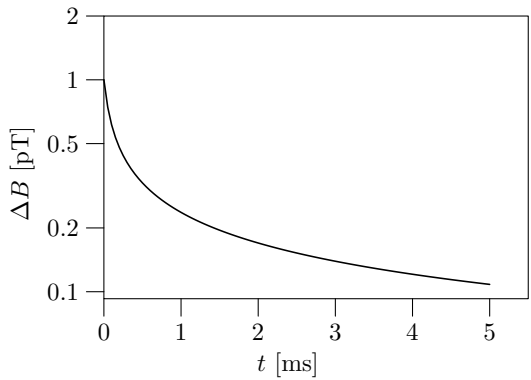}
  \caption{Uncertainty of two $B$ field components as a function of time. We
    use a single atomic gas and two probe beams which are turned on
    simultaneously. We have chosen a segment duration $\tau =
    10^{-8}$~s and corresponding field parameters $\kappa_\tau^2 =
    0.0183$ and $\mu_\tau = 8.8 \times 10^{-4}$. The joint uncertainty
    of the two $B$ fields at $t = {\unit[5]{\milli\second}}$ is
    $\Delta B_y = \Delta B_z = {\unit[0.1083]{\pico\tesla}}$. }
  \label{fig:3}
\end{figure}
Thus effectively we have no squeezing of the atoms leaving us with
the $\nicefrac{1}{t}$ decrease in the variance of the $B$
components as shown in Fig.~\ref{fig:3}.

To estimate simultaneously and precisely two $B$ field components
we need to measure two commuting atomic variables. Such a
measurement is possible by using two separate gasses and by
estimating one $B$ field component on each system with individual
probe beams. Both systems are then equivalent to the setup in
Fig.~\ref{fig:5} used to measure $B_y$, but in one system (not
shown), the probe beam propagates along the $z$ axis such that we
estimate $B_z$ by measuring $p_\mathrm{ph}$. The result is similar
to the result in Fig.~\ref{fig:1}
except that the number of atoms and the photon flux are both divided
by two as they are shared between the two systems. The uncertainty of
the $B$ fields for large $t$ is proportional to
$\nicefrac{1}{\sqrt{N_\mathrm{at}^2\Phi t^3}}$ so the uncertainty for
estimating two $B$ field components is $2\sqrt{2}$ times larger than
if we had measured only one with the same field and atomic resources.

The achievements of a sequential measurement of first $B_y$ and then
$B_ z$ are shown in Fig.~\ref{fig:4}. In the first half of the time
the full photon flux is spent to measure $B_ y$, leaving the
uncertainty about $B_ z$ unchanged. Although $B_ z$ is subsequently
coupled to an anti-squeezed atomic component, it is quickly squeezed,
and the estimate of $B_z$ shows a
$\nicefrac{1}{\sqrt{N_\mathrm{at}^2\Phi t^3}}$ dependence as in
Eq.~\eqref{eq:21}. Now the other atomic observable is anti-squeezed,
but this will surely not degrade our information already obtained
about the classical $B_y$ component. The values of $\var(B_y(t))$ and
$\var(B_z(t))$ are $2\sqrt{2}$ times larger than the uncertainty reported
in Fig.~\ref{fig:1} because only half the time is spent on the
measurement of each component.
\begin{figure}[htbp]
  \centering
  \includegraphics[width=7cm]{figvivi4.eps}
  \caption{Uncertainty of two $B$ field components as a function of
    time using one atomic gas. $B_y$ is estimated in the first half of
    the time and $B_z$ in the second half. The dashed line is for
    $B_y$ and the value at $t={\unit[5]{\milli\second}}$ is $\Delta
    B_y = {\unit[1.647\times 10^{-4}]{\pico\tesla}}$.  The full line
    is for $B_z$ and the value at $t={\unit[5]{\milli\second}}$ is
    $\Delta B_z = {\unit[1.645\times 10^{-4}]{\pico\tesla}}$. We have
    chosen a segment duration $\tau = 10^{-8}$~s and corresponding
    field parameters $\kappa_\tau^2 = 0.0183$ and $\mu_\tau = 8.8
    \times 10^{-4}$.}
  \label{fig:4}
\end{figure}

\subsection{Two components: two entangled gasses and two probe beams}
\label{sec:two-entangled-gasses}

\begin{figure}[htbp]
  \centering
  \includegraphics{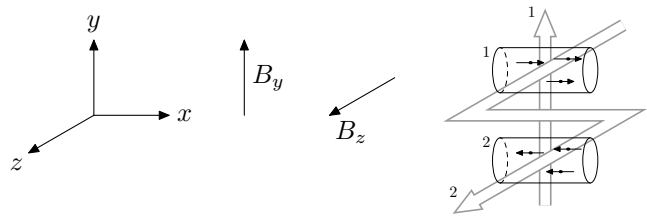}
  \caption{Setup for measuring two components of a $B$ field using two
    entangled gasses and two probe beams.}
  \label{fig:6}
\end{figure}

If we split the atomic sample into two and polarize one gas along
$x$ and the other along $-x$ such that $\langle J_{x_1}\rangle =
-\langle J_{x_2}\rangle$, then the two observables $(J_{y_1}-J_{y_
2})$ and $(J_{z_ 1}- J_{z_ 2})$, and equivalently
$x_{\mathrm{at}_1} - x_{\mathrm{at}_2}$ and $p_{\mathrm{at}_1} -
p_{\mathrm{at}_2}$ commute. We couple these observables to the $B$
fields and probe beams and use the setup shown in Fig.~\ref{fig:6}
(see also the entanglement experiment of
Ref.~\cite{julsgaard01:_exper_long_entan_two_macros_objec}).

Both optical probe beams have $S_x$ classical, and one beam
propagates along $y$, the other along $z$.
The beams pass through both gasses and, oppositely to the protocol
in the previous section, we use all atoms to estimate the $B$
field components. The effective Hamiltonian for the setup is
\begin{gather}
  \label{eq:24}
  \begin{split}
    \mathcal{H}_\mathrm{int}\tau &= \mu_\tau B_y(x_{\mathrm{at}_1} +
    x_{\mathrm{at}_2}) + \mu_\tau B_z(p_{\mathrm{at}_1} +
    p_{\mathrm{at}_2})\\
    &\quad + \kappa_\tau(p_{\mathrm{at}_1} -
    p_{\mathrm{at}_2})p_{\mathrm{ph}_1} + \kappa_\tau(x_{\mathrm{at}_1} -
    x_{\mathrm{at}_2})x_{\mathrm{ph}_2},
  \end{split}
\end{gather}
where the two minus signs can be implemented by changing the sign
on $\kappa_\tau$ after the probe beams have passed through the
first gas. In practice the change in sign can be effectuated by
changing the sign of the detuning or by interchanging $\sigma^+$
and $\sigma^-$ polarizations with, e.g., a
half-wave-plate~\cite{shurcliff62:_polar_light}. The gaussian
state vector is $\mathbf{y} = (B_z, B_y, x_{\mathrm{at}_1},
p_{\mathrm{at}_1}, x_{\mathrm{at}_2}, p_{\mathrm{at}_2},
x_{\mathrm{ph}_1}, p_{\mathrm{ph}_1}, x_{\mathrm{ph}_2},
p_{\mathrm{ph}_2})^T$ and from the Heisenberg equations of motion
for the operators we get the following transformation matrix
\begin{gather}
  \label{eq:25}
  \mathbf{S}_\tau =
  \begin{pmatrix}
    1 & 0 & 0 & 0 & 0 & 0 & 0 & 0 & 0 & 0\\
    0 & 1 & 0 & 0 & 0 & 0 & 0 & 0 & 0 & 0\\
    \mu & 0 & 1 & 0 & 0 & 0 & 0 & \kappa & 0 & 0\\
    0 & -\mu & 0 & 1 & 0 & 0 & 0 & 0 & -\kappa & 0\\
    -\mu & 0 & 0 & 0 & 1 & 0 & 0 & \kappa & 0 & 0\\
    0 & \mu & 0 & 0 & 0 & 1 & 0 & 0 & -\kappa & 0\\
    0 & 0 & 0 & \kappa & 0 & -\kappa & 1 & 0 & 0 & 0\\
    0 & 0 & 0 & 0 & 0 & 0 & 0 & 1 & 0 & 0\\
    0 & 0 & 0 & 0 & 0 & 0 & 0 & 0 & 1 & 0\\
    0 & 0 & -\kappa & 0 & \kappa & 0 & 0 & 0 & 0 & 1
  \end{pmatrix}.
\end{gather}

\begin{figure}[htbp]
  \centering
  \includegraphics[width=7cm]{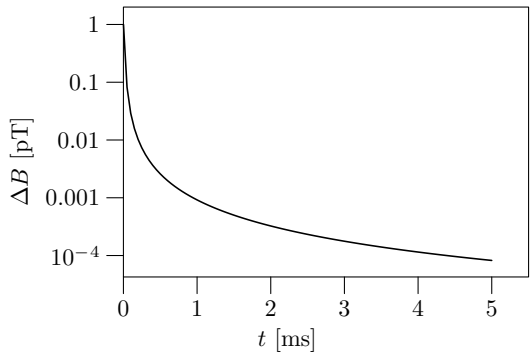}
  \caption{Uncertainty of two $B$ field components as a function of
    time using two atomic entangled gasses and two probe beams. We
    have chosen a segment duration $\tau = 10^{-8}$~s and
    corresponding field parameters $\kappa_\tau^2 = 0.0183$ and
    $\mu_\tau = 8.8 \times 10^{-4}$.  The uncertainty at $t =
    {\unit[5]{\milli\second}}$ is $\Delta B_y = \Delta B_z =
    {\unit[8.221\times 10^{-5}]{\pico\tesla}}$.  }
  \label{fig:15}
\end{figure}

The time evolution of the uncertainty of the $B$ fields is shown
in Fig.~\ref{fig:15}. The final uncertainty of $B_y$ and $B_z$ is a
factor $\sqrt{2}$ higher than in Fig.~\ref{fig:1},
where we used the entire photon flux to probe only a single $B$ field
component, but a factor of two lower than
in the setup with separate probing of non-entangled gasses. These
factors were
expected because of the $\nicefrac{1}{\sqrt{N_\mathrm{at}^2\Phi t^3}}$
dependence of Eq.~\eqref{eq:21}.
By using entangled gasses we use all atoms, but only half of the
photon flux to estimate each $B$ field component.

\subsection{Three dimensional vector magnetometry}
\label{sec:three-separ-gass}

\begin{figure}[htbp]
  \centering
  \includegraphics{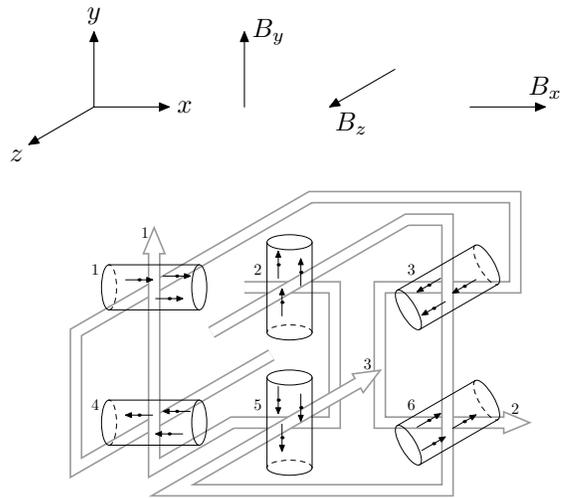}
  \caption{Setup used to obtain estimates of all three components of a
    magnetic field. We use six atomic gasses and three probe
    beams.}
  \label{fig:7}
\end{figure}
For the estimation of three components of a magnetic field,  the
situation changes since a spin polarized sample is not a probe for
the field component parallel with the spin. For the sequential
probing one would thus measure $B_ y$ and $B_ z$ as just
described, but one would have to rotate the sample by $90^\circ$
to determine the last component $B_ x$, and the errors of such a
rotation will limit the precision.

As an alternative, we can divide the gas into three samples which are
spin polarized along different directions. With such a setup, we
can perform independent estimates of the three components. Since each
component is determined by one third of the atoms and one third of
the photons the scaling of Eq.~\eqref{eq:21}
predicts that the uncertainty
is a factor of $3\sqrt{3}$ larger than in Fig.~\ref{fig:1} where
we only estimated one $B$ field component.

We wish to couple the $B$ field components to
three \textit{commuting} atomic operators, involving as many atoms
as possible, and a better, but also more complicated, setup is shown in
Fig.~\ref{fig:7}. Here we use six entangled gasses and we
let three probe beams pass through four gasses each. The gasses
are polarized with the following macroscopic components: $\langle
J_{x_1}\rangle = -\langle J_{x_4}\rangle$, $\langle J_{y_2}\rangle
= -\langle J_{y_5}\rangle$, and $\langle J_{z_3}\rangle = -\langle
J_{z_6}\rangle$. The optical fields are linearly polarized with
macroscopic Stokes parameters $\langle S_{z_1}\rangle$, $\langle
S_{y_2}\rangle$, and $\langle S_{x_3}\rangle$. The Hamiltonian for
the system shown in Fig.~\ref{fig:7} is given by
\begin{gather}
  \label{eq:2}
  \begin{split}
    \mathcal{H}_\mathrm{int}\tau \sqrt{\hbar \langle J_{x_1} \rangle}
    &=
    \mu_\tau(J_{x_2}+J_{x_3}+J_{x_5}+J_{x_6})B_x\\
    &\quad + \mu_\tau(J_{y_1}+J_{y_3}+J_{y_4}+J_{y_6})B_y\\
    &\quad + \mu_\tau(J_{z_1}+J_{z_2}+J_{z_4}+J_{z_5})B_z\\
    &\quad + \kappa_\tau(J_{z_2}-J_{y_3}-J_{z_5}+J_{y_6})S_3\\
    &\quad + \kappa_\tau(J_{z_1}-J_{x_3}-J_{z_4}+J_{x_6})S_2\\
    &\quad + \kappa_\tau(J_{y_1}-J_{x_2}-J_{y_4}+J_{x_5})S_1,
  \end{split}
\end{gather}

Terms like $\mu_\tau J_{x_1}B_x$, which couple the classical
components of the atomic spins to the $B$ fields, are omitted from
the interaction Hamiltonian as they do not contribute to the
interactions to the same order, e.g., $[J_ {y_1}, \mu_\tau J_{x_1}
B_x ] = - i \mu_\tau \hbar J_{z_1} B_ x$, the product of two small
quantities. By using the Hamiltonian of Eq.~\eqref{eq:2} we
measure three commuting observables
$J_{z_2}-J_{y_3}-J_{z_5}+J_{y_6}$,
$J_{z_1}-J_{x_3}-J_{z_4}+J_{x_6}$, and
$J_{y_1}-J_{x_2}-J_{y_4}+J_{x_5}$ and from their commutators with
the Larmor term in Eq.~\eqref{eq:2}, we see that they evolve in direct
proportion with the three $B$ field components.
\begin{figure}[htbp]
  \centering
  \includegraphics[width=7cm]{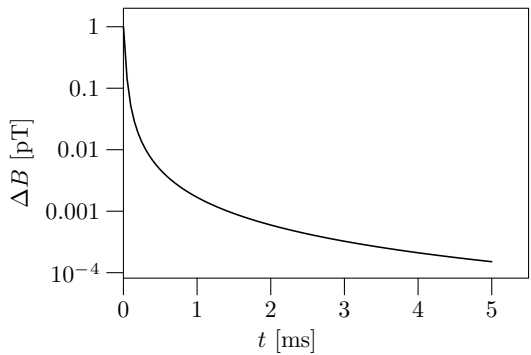}
  \caption{Uncertainty of three $B$ field components using six entangled
    gasses and three probe beams. We have chosen a
    segment duration $\tau = 10^{-8}$~s and corresponding field
    parameters $\kappa_\tau^2 = 0.0183$ and
    $\mu_\tau = 8.8 \times 10^{-4}$. The uncertainties at
    $t={\unit[5]{\milli\second}}$ are $\Delta B_x = \Delta B_y =
    \Delta B_z = {\unit[1.510\times 10^{-4}]{\pico\tesla}}$.
    }
  \label{fig:17}
\end{figure}

The uncertainty of the $B$ field components is shown as a function of
time in Fig.~\ref{fig:17}. As we use $\nicefrac{4}{6}$ of the atoms
and $\nicefrac{1}{3}$ of the photon flux to estimate each $B$ field
component, the value of the uncertainty at
$t={\unit[5]{\milli\second}}$ is $\nicefrac{3\sqrt{3}}{2}$ times
larger than if we use all atoms and all photons to estimate only one
$B$ field component and a factor of two smaller than if we use three
separate systems to estimate the three $B$ field components.

\section{Quantifying entanglement between atomic samples}
\label{sec:entanglement}

To measure two or three $B$ field components most efficiently, we
showed that one should use entangled gasses. Here we quantify the
degree of entanglement by calculating the gaussian entanglement of
formation (GEoF)~\cite{giedke03:_entan_format_symmet_gauss_states}
from the covariance matrix of the atoms
$\boldsymbol{\gamma}_\mathrm{atoms}$. Every time we apply the update
formula, we may extract $\boldsymbol{\gamma}_\mathrm{atoms}$ for a
pair of gasses from our numerical procedure and up to local unitary
operations this matrix turns out to be on the form
\begin{gather}
  \label{eq:38}
  \boldsymbol{\gamma}_\mathrm{atoms} =
  \begin{pmatrix}
    n & 0 & k_x & 0\\
    0 & n & 0 & -k_p\\
    k_x & 0 & n & 0\\
    0 & -k_p & 0 & n
  \end{pmatrix}
\end{gather}
where $n$, $k_x$, and $k_p$  ($k_x = k_p$, in our case) are the
quantities of interest for the evaluation of the GEoF,
$  E = c_+(\Delta) \log[c_+(\Delta)] - c_-(\Delta)
\log[c_-(\Delta)]$,
with $c_\pm(\Delta) = \frac{1}{4} (\Delta^{-\nicefrac{1}{2}} \pm
\Delta^{\nicefrac{1}{2}})^2$, and $\Delta =
\min\left(1,\sqrt{(n-k_x)(n-k_p)}\right)$.

\begin{figure}[htbp]
  \centering
  \includegraphics{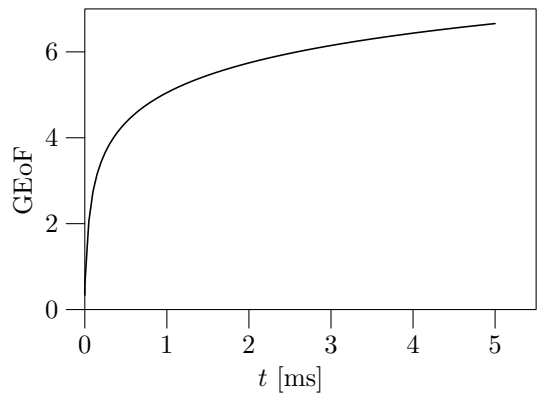}
  \caption{GEoF for two entangled gasses corresponding to the case
    considered in Fig.~\ref{fig:15}.}
  \label{fig:19}
\end{figure}

The GEoF for the two gasses used to estimate two $B$ field
components is shown in Fig.~\ref{fig:19}.
\begin{figure}[htbp]
  \centering \includegraphics{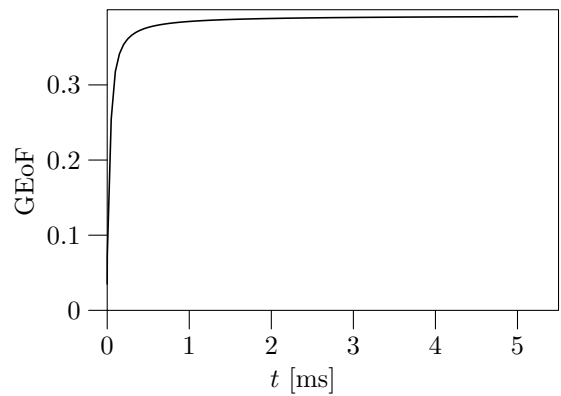}
  \caption{GEoF for two gasses polarized in opposite directions as
    considered in Fig.~\ref{fig:17}.}
  \label{fig:20}
\end{figure}
For three $B$ field components, we used six atomic gasses, and we
have calculated the GEoF between different pairs of the gasses.
Figure~\ref{fig:20} shows the GEoF between two gasses polarized in
opposite directions, e.g., gas number~1 and 4 in Fig.~\ref{fig:17}. The GEoF
between pairs like 1 and 2 is zero.

The setup with two gasses is quite equivalent to the one implemented
in recent entanglement
experiments~\cite{julsgaard01:_exper_long_entan_two_macros_objec}
except that the atomic systems are under the additional influence of
an initially unknown $B$ field. This slows down the initial rate of
generation of entanglement, but as $\var(B(t))$ approaches zero, the
entanglement grows without limits as long as absorption and atomic
decay can be neglected~\cite{sherson}. In the case of six gasses
which are probed in a non-symmetric way, some pairs show entanglement
and some do not. This can be understood by identification of operators
that do not couple to the probe fields. The convergence of the
entanglement between oppositely polarized gasses towards a constant
value is also observed without coupling to a $B$ field, and it is due
to the incompleteness of the measurements on the pair by fields that
also couple to other pairs of gasses. In symmetric setups with
multiple gaussian variables, the theoretical maximum of pairwise
entanglement between systems also have upper limits reflecting the
impossibility for a quantum system to be maximally entangled with
several other quantum systems at the same
time~\cite{wolf04:_ent_fru_gaus,plenio04:_entropy}.

\section{Effects of noise}
\label{sec:noise}

Effects of noise were recently discussed in
Ref.~\cite{auzinsh04:_can_a}. Here we include noise in our
gaussian description. As the photonic probe beams pass through the
atomic sample, there is a probability for photon
absorption~\cite{hammerer:_light_matter_quant_inter}
$  \epsilon = N_\mathrm{at} \frac{\sigma}{A}
  \frac{\nicefrac{\Gamma^2}{4}}{ \nicefrac{\Gamma^2}{4} +
  \Delta^2}$
and a related probability of atomic decay~\cite{hammerer:_light_matter_quant_inter,molmer04:_estim_class_param_gauss_probes}
$  \eta_\tau = \Phi \tau \frac{\sigma}{A}
\frac{\nicefrac{\Gamma^2}{4}}{
  \nicefrac{\Gamma^2}{4} + \Delta^2}$,
where $\Gamma$ is the atomic decay rate, $\sigma =
\nicefrac{\lambda^2}{(2\pi)}$ is the resonant photon absorption
cross-section, $A$ is the beam cross-section, and $\Delta$ the
detuning. These processes lead to a reduction in the polarization of
the Stokes vector and the atomic spin and to an incoherent noise
contribution. These features were discussed in
Refs.~\cite{hammerer:_light_matter_quant_inter,molmer04:_estim_class_param_gauss_probes}
and at length in Ref.~\cite{madsen04:_spin_squeez_precis_probin_light}
so here it is sufficient to recall the generalizations of the update
formulae in Eqs.~\eqref{eq:11}--\eqref{eq:28}:
\begin{align}
  \label{eq:8}
  \mathbf{m}(t+\tau) &= \mathbf{L}_\tau \mathbf{S}_\tau \mathbf{m}(t)\\
  \label{eq:31}
  \boldsymbol{\gamma}(t+\tau) &= \mathbf{L}_\tau \mathbf{S}_\tau
  \boldsymbol{\gamma}(t) \mathbf{S}_\tau^T \mathbf{L}_\tau +
  \frac{\hbar N_\mathrm{at}}{ \langle J_x(t)\rangle} \mathbf{M}_\tau +
  \frac{\hbar N_\mathrm{ph}}{ 2\langle S_x(t)\rangle} \mathbf{N},
\end{align}
where the diagonal matrices $\mathbf{L}_\tau = \diag(1,
\sqrt{1-\eta_\tau}, \sqrt{1-\eta_\tau}, \sqrt{1-\epsilon},
\sqrt{1-\epsilon})$, $\mathbf{M}_\tau =
\diag(0,\eta_\tau,\eta_\tau,0,0)$ and $\mathbf{N}=
\diag(0,0,0,\epsilon,\epsilon)$ describe the loss of polarization, and
the noise introduced by stimulated emission and photon absorption,
respectively. In each time step $\tau$ the atomic polarization
$\langle J_x \rangle$ is reduced by the factor $(1-\eta_ \tau)$ and
$\kappa_\tau$ is reduced by $\sqrt{1-\eta_\tau}$. The
presence of noise leaves the update formula for measurements in
Eq.~\eqref{eq:27} unchanged. We may analyze the pertaining Ricatti
equation, which with $\mathbf{y} = (B, p_\mathrm{at})^T$ can be
written on the same form as in Eq.~\eqref{eq:20} but now
with $\mathbf{C} = \left(
\begin{smallmatrix}
  0 & 0\\
  0 & \frac{N_\mathrm{at}}{\langle J_x(t)\rangle} \eta
\end{smallmatrix}
\right)$,
$ \mathbf{D} = \left(
\begin{smallmatrix}
  0 & 0\\
  \mu & \frac{\eta}{2}
\end{smallmatrix}
\right)$,
$\mathbf{E} = \mathbf{D}^T$, and $\mathbf{B} = \left(
\begin{smallmatrix}
  0 & 0\\
  0 & \frac{(1-\epsilon)\kappa^2(t)}{1 - \epsilon\left(1 -
  \frac{N_\mathrm{ph}}{2\langle S_x(t)\rangle}\right)}
\end{smallmatrix}
\right)$, where $\eta = \eta_\tau/\tau$. One can show that the noise
terms in $\mathbf{D}$ and $\mathbf{E}$ will have vanishing effect, and
if we restrict ourselves to times corresponding to $\eta t \ll 1$, we
may neglect the time dependence of $\langle J_x\rangle$, $\langle
S_x\rangle$, and $\kappa$ such that $\nicefrac{N_\mathrm{at}}{\langle J_x\rangle} =
2$, $\nicefrac{N_\mathrm{ph}}{\langle S_x\rangle} = 2$, and $\kappa(t)
= \kappa(0)$. With these
approximations, the resulting linear equations for the matrices
$\mathbf{U}$ and $\mathbf{W}$ can be solved analytically, leading to
lengthy expressions with sums of products of exponential functions,
constant terms, and
terms linear in time $t$. In the limit $\sqrt{\eta\kappa^2}t \gg 1$,
it is an accurate approximation to maintain only the leading
exponential and we find
\begin{gather}
  \label{eq:1}
  \var(B(t)) \to \frac{\eta}{\mu^2t}.
\end{gather}
Compared with the result in Eq.~\eqref{eq:21}, we note that in the
long time limit the uncertainty decreases as
$\nicefrac{1}{N_\mathrm{at}t}$, and not as
$\nicefrac{1}{N_\mathrm{at}^2t^3}$ as in the noise-less case.

Figure~\ref{fig:0} shows (solid line) the time evolution of the
uncertainty of the $B$ field. For short times the analytical
expression~\eqref{eq:21} disregarding the noise (dotted line)
in the figure, is a good approximation, and for long times the
numerical result follows the expression~\eqref{eq:1}, shown as a
dashed curve on the interval between ${\unit[0.5]{\milli\second}}$ and
${\unit[5]{\milli\second}}$. Similar results are obtained in the
simulation of measurements of two and three $B$ field components.
\begin{figure}[htbp]
  \centering
  \includegraphics[width=7cm]{figvivi11.eps}
  \caption{Uncertainty of one $B$ field component as a function of
    time. The full line is the result of a full numerical calculation
    and the value at $t={\unit[5]{\milli\second}}$ is $\Delta B_y =
    {\unit[2.333\times 10^{-4}]{\pico\tesla}}$. The dotted line is
    without inclusion of noise and the value at
    $t={\unit[5]{\milli\second}}$ is $\Delta B_y =
    {\unit[5.814\times10^{-5}]{\pico\tesla}}$. We have chosen a
    segment duration $\tau = 10^{-8}$~s and corresponding field and
    noise parameters $\kappa_\tau^2 = 0.0183$, $\mu_\tau = 8.8 \times
    10^{-4}$, $\eta_\tau = 1.76 \times 10^{-8}$ and $\epsilon =
    0.0281$. The dashed curve shows the result of Eq.~\eqref{eq:1}
    valid for $t \ll {\unit[1]{\second}}$, and $\sqrt{\eta\kappa^2t}
    \gg 1$.}
  \label{fig:0}
\end{figure}

\section{Conclusion and Outlook}
\label{sec:conclusion-outlook}

We have considered how to estimate a vector magnetic field using a
gaussian description of the variables describing the system. To
estimate more than one field component it is fruitful to use pairwise
entangled separate gasses. In other applications, e.g.\ teleportation,
shared entanglement over some distance is a useful resource. We showed
that entanglement can be a useful local resource to improve the
accuracy in measurements and parameter estimation. We note that our
protocol for the estimation of two $B$ field components using two
optical probe beams and two entangled gasses is experimentally very
feasible. Essentially, it would require a combination of the
magnetometry setup of Ref.~\cite{geremia04:_sub-shotnoise} and the
entanglement setup of
Ref.~\cite{julsgaard01:_exper_long_entan_two_macros_objec}.

In a broader perspective, the present work brings out the virtues
of the gaussian state formalism when it comes not only to the
detailed characterization of entanglement (see, e.g.,
Refs.~\cite{giedke03:_entan_trans_pure_gauss_states,eisert03:_introd_basic_entan_theor_contin_variab_system,fiurasek02:_gauss_trans_distil_entan_gauss_states,giedke02:_charac_gauss_operat_distil_gauss_states} and references
therein), but also to the practical description of quantum systems~\cite{hammerer:_light_matter_quant_inter,molmer04:_estim_class_param_gauss_probes,madsen04:_spin_squeez_precis_probin_light}.
The fact that the gaussian state is fully characterized in terms
of its expectation value vector and covariance matrix means that
the theory is easy to formulate and to evaluate. It is an
outstanding advantage of the gaussian state description that
explicit update formulae exist not only for the interaction dynamics but also
for the back-action due to measurement. The gaussian state
description is a very versatile tool and we foresee this approach
to be used for the description of a variety of different systems
in the near future.

\begin{acknowledgments}
L.B.M. is supported by the Danish Natural Science Research Council
(Grant No. 21-03-0163).
\end{acknowledgments}


\end{document}